\documentclass[aps,prb,twocolumn]{revtex4-1}
\usepackage[english]{babel}
\usepackage[utf8]{inputenc}
\usepackage{amsmath}
\usepackage{amssymb}
\usepackage[caption=false]{subfig}
\usepackage{amssymb}
\usepackage{epsfig}
\usepackage{graphicx}
\usepackage{amsmath}
\usepackage{array,color}
\usepackage{natbib}
\usepackage{soul}
\usepackage[usenames,dvipsnames]{xcolor}
\usepackage{color, colortbl}
\definecolor{forestgreen}{rgb}{0.11,0.54,0.15}
\definecolor{purple}{rgb}{0.62,0.10,0.96}
\definecolor{dockerblue}{rgb}{0.11,0.56,0.98}
\definecolor{freeblue}{rgb}{0.25,0.41,0.88}
\usepackage[pdftex,plainpages=false,colorlinks=true,linkcolor=Red, citecolor=blue, urlcolor=blue]{hyperref}

\newcommand*{\bfrac}[2]{\genfrac{}{}{0pt}{}{#1}{#2}}

\begin{document}
\title{Doping-driven pseudogap-metal-to-metal transition in correlated electron systems}
\author{L. Fratino$^{1}$, S. Bag$^{1}$, A. Camjayi$^{2}$, M. Civelli$^{1}$, M. Rozenberg$^{1}$}
\affiliation{{$^1$ Universit\'e  Paris-Saclay, CNRS Laboratoire de Physique des Solides, 91405, Orsay, France}\\
{$^2$ Departamento de F\'isica,FCEyN, UBA and IFIBA, Conicet, Pabell\'on 1, Ciudad Universitaria, Buenos Aires 1428 CABA, Argentina}}

\date{\today}

\begin{abstract}
  We establish that a doping-driven first-order metal-to-metal transition, from a pseudogap metal to Fermi Liquid, can occur
  in correlated quantum materials. Our result is based on the exact Dynamical Mean Field Theory solution of the Dimer Hubbard Model.
  This transition 
  elucidates the origin of many exotic features in doped Mott materials, like the pseudogap in cuprates,
  incoherent bad metals, enhanced compressibility and orbital selective Mott transition.
  This phenomenon is suggestive to be at the roots of the
  many exotic phases 
  appearing in the phase diagram of correlated materials.

\end{abstract} 
\maketitle
Electronic correlations put into question our current understanding of metals.
Emblematic is the case of Mott insulators\cite{RevModPhys.40.677, RevModPhys.70.1039, PhillipsRMP:2010}
(like e.g. transition metal oxides, Bechgaard organic salts, alkali-doped fullerides, twisted bilayer graphene, etc).
According to the standard Bloch band-theory, these
materials should be metals, but they turn out to be robust insulators because of
dominating electronic correlations \cite{ANDERSON1196,varma1987charge}. 
Upon doping, a metallic state is recovered, but
often displaying unconventional properties. 
One of the most striking examples is the so called pseudogap (PG) metal\cite{warren1989cu,alloul198989}
that appears for small doping in the phase diagram of cuprates
superconductors, and which remains still poorly understood \cite{normanADV}.
A conventional metallic phase, which can be described within the conventional Fermi Liquid (FL) Theory,
is reestablished only for high doping \cite{Peets_2007}.

How the correlated Mott insulator evolves through those ``bad metallic'' states into the FL as a
function of doping has remained a key open issue, not only in the context of cuprates.
A point of view, that can be ascribed to the early work of P.W. Anderson \cite{ANDERSON1196},
advocates the existence of a new kind of metallic state,
characterized by the absence of any broken symmetry even at the
lowest temperatures.
In this state, carriers have to move through a backround described as a liquid of dynamical singlets, the RVB state. 
Another point of view advocates the existence of a quantum phase transition between
an exotic order-parameter metal and the conventional FL \cite{https://doi.org/10.1002/pssb.200983037,Varma_2016}.
Unconventional phenomena, like the PG in cuprates, would then be the byproduct of this order parameter.
In most cases, however, it has been difficult or impossible to identify such order, and
the issue has remained openly debated (see e.g. the aforementioned PG phase or the so called
hidden order in  URu$_2$Si$_2$ \cite{PhysRevLett.55.2727,Mydosh_2020}).
Recent proposals have attempted to reconciles these two scenarios
by introducing the concept of inter-twinning \cite{RevModPhys.87.457,Keimer2015},
or competition \cite{Efetov2013,PhysRevLett.111.027202} of different orders to form
a new un-ordered unconventional phase, akin to Anderson's proposal\cite{ANDERSON1196}.
Establishing the existence 
of the PG-metal-to-metal transition is therefore a fundamental
step within this open debate that 
could allow us to finally 
unveil the origin of key phenomena in correlated electron physics,
like the high temperature superconductivity.

Recent works
\cite{PhysRevLett.104.226402,PhysRevB.84.075161,PhysRevLett.108.216401,fratino2016organizing,PhysRevLett.120.067002,Cha18341,PhysRevB.99.054516},
within the context of the two-dimensional Hubbard-like Models studied
with state of the art methods like cluster extensions of the
Dynamical Mean Field Theory (DMFT)
have proposed the existence of a first order quantum phase
transition between a PG and a FL metals as a function of doping.
In two dimensions, however, 
only approximate solutions can be provided. 
Unfortunately, issues such as the "minus sign problem" prevent those state-of-the-art methods 
to obtain
complete phase diagrams as a function of interaction, chemical potential and temperature \cite{PhysRevX.11.011058}.
A consensus on the existence of the metal-to-metal transition has not been achieved yet.
 
Here we provide a concrete answer to this problem by an exact DMFT solution  of a minimal model Hamiltonian,
which establishes the existence of
the first-order PG metal to FL metal transition.
We focus on the dimer Hubbard Model (DHM), which
 a priori has the necessary ingredients to capture the relevant physics,
 namely, strong local Coulomb repulsion and 
 non-local magnetic exchange interactions. 
 The latter is a key feature, which is missing in the 
 single-site Hubbard model, but is present in the 2D models mentioned before. 
 However, in contrast to the 2D models, the DHM has the significant merit that it can 
 be exactly solved within DMFT  \cite{PhysRevB.59.6846,PhysRevB.95.035113}.
Our results partially reconcile the debate between PG-metal vs 
quantum phase transition scenarios.
We find that upon doping the DHM Mott state,
an unconventional bad metal appears, displaying PG features reminiscent of the PG in cuprates and possessing a
dominant spin-singlet component, akin to Anderson's proposal \cite{ANDERSON1196}. Differently from the latter,
however, a first order quantum phase transition to
a more conventional FL does take place by further increase of doping.
This PG-metal-to-metal transition, however, does not need to advocate for an
elusive order parameter, but appears to be of the liquid-vapor type
\cite{PhysRevLett.104.226402,PhysRevB.84.075161,PhysRevLett.108.216401,fratino2016organizing}.

Similarly to the case of the Mott metal-insulator transition in the single site Hubbard model,
our present results suggest that the PG-metal-to-metal transition in the DHM may have universal character within 
correlated quantum materials.
As we shall see, many unconventional properties of correlated materials are naturally realized in the
solution of the model, such as a pseudogap, incoherent or bad metallicity, enhanced compressibility, 
and orbital selectivity. 
Thus, we may argue that the present  PG-metal-to-metal transition 
may be the long-sought phenomenon where many exotic quantum states 
find their roots \cite{Keimer2015}.

\begin{figure}[!ht]
\begin{center}
\includegraphics[width=.9\linewidth]{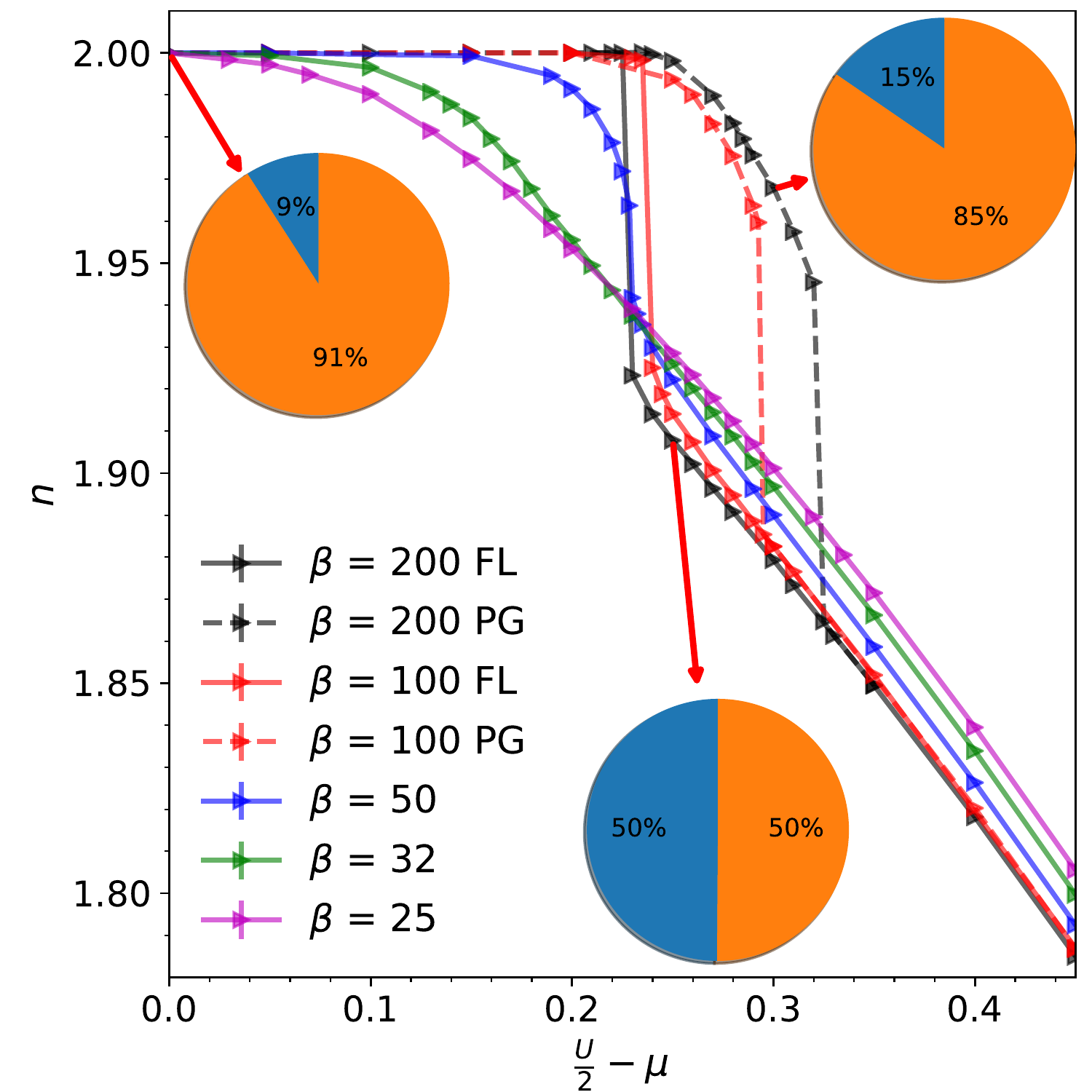}
\end{center}
\caption{The electron density $n$ as function of the renormalized chemical potential $\frac{U}{2}-\mu$. 
Half-filling corresponds to $n = 2$, where the system is a Mott insulator.
  The hysteresis behavior below $\beta = 50$ defines a PG-metal-to-metal coexistence region. 
  Note the enhancement of the compressibility 
  at the endpoint of the first order transition (blue line) at $\beta = 50$.
 The orange sector of the pie-charts indicates the relative
 contribution of the singlet states to the DAIM wave-function projected on the 
 isolated dimer. The blue sector corresponds to all other states (see details in the SM \cite{sup1}).}
  \label{nVsmu}
\end{figure}
 The Dimer Hubbard Model (DHM) consists of a lattice of dimers:
\begin{align}\label{HH}
 {\mathcal{H}} = \sum_{i, j; \sigma= \uparrow, \downarrow}  {\psi}^{\dagger}_{i,\sigma} \, \mathbf{T}_{i,j} \, {\psi}_{j,\sigma}
  + U \sum_{i;\alpha=1,2} n_{i\alpha\uparrow} \, n_{i\alpha\downarrow} 
\end{align}
The spinor ${\psi}_{i,\sigma}=\, ( c_{i,1,\sigma}, c_{i,2,\sigma} )$ acts on the dimer sites $\alpha=1,2$ at the lattice site $i$,
being $c^{\dagger}_{i\alpha\sigma}$, $c_{i\alpha\sigma}$ the electron creation and destruction operators respectively.
The matrix $\mathbf{T}_{i,j}=\, -t \hat{1}$ and $\mathbf{T}_{i,i}=\, -\mu \hat{1}+ \, t_\perp \hat{\sigma}_x$
($\hat{\sigma}_x$ is the first Pauli matrix)
describes the nearest-neighbor inter-dimer and the intra-dimer electron hopping respectively.
Electrons experience strong correlation via the on-site local repulsion $U$. 
We adopt $t=0.5$ and $t_\perp=-0.3$, and 
fix the on-site local repulsion term $U= 2.3$ in order to be deep in the correlated regime.
The same set of parameters has previously been considered in the study of the model
at half-filling, where it exhibits a first-order 
temperature-driven insulator-to-metal transition \cite{PhysRevB.95.035113,PhysRevB.97.045108}.
Here, we shall also vary the chemical potential $\mu$ to induce a doping-driven 
insulator-to-metal transition by adding or removing electrons.


The advantage of the DHM is that in the infinite dimensional limit DMFT provides the exact solution \cite{georges2,PhysRevB.59.6846,PhysRevB.73.245118, PhysRevB.75.193103,hafermann2009metal,PhysRevB.95.035113,PhysRevB.97.045108}. We can then establish
properties of the doped metallic state and its possible phase transitions without the uncertainty introduced by 
an approximation, such as in the CDMFT treatments of the 2D Hubbard model on a square lattice. 
It is convenient to express the DMFT equations
in the diagonal bonding/anti-bonding ($B/AB$) basis of the lattice hopping matrix $\mathbf{T}_{i,j}$.
The dimer Green's function in the $B/AB$ basis can be written in
terms the components of site basis Green's function $G_{\alpha \beta}$ as
diag$({G_{B}, G_{AB}})$ = diag$({G}_{11} - {G}_{12}, {G}_{11} + {G}_{12})$, 
where we dropped the spin indices for clarity. 
In the infinite dimensional limit, adopting a semicircular (i.e. Bethe lattice) 
density of states $D(\epsilon) =-\frac{1}{t \pi}\sqrt{(1- (\frac{\epsilon}{2t})^2)}$,
the DMFT self-consistency equations are then readily written:
\begin{align}\label{selfcon1}
{{\cal G}_o}^{-1}_{B/AB} (\omega)= \omega+\mu \pm t_\perp - t^2 G_{B/AB}(\omega)
\end{align}
where ${{\cal G}_o}_{B/AB} (\omega)$ is the Weiss field describing the bath of the dimer Anderson impurity model (DAIM)
associated to the Hamiltonian (\ref{HH}). We solve the DAIM imaginary-time action 
\begin{flalign}\label{selfcon2}
  S_{\rm DAIM}= & -\int_0^\beta d\tau d\tau^\prime \sum_{ \bfrac{\sigma= \uparrow,\downarrow} {\alpha, \beta= 1,2}} 
  c^{\dagger}_{\alpha\sigma}(\tau) \, {{\cal G}_o}^{-1}_{\alpha\sigma,\beta\sigma}(\tau-\tau^\prime)
  \, c_{\beta\sigma}(\tau^{\prime})+ &  \nonumber \\
         & +U \int_0^\beta d\tau \sum_{\alpha= 1,2}\, n_{\alpha\uparrow}(\tau) n_{\alpha\downarrow}(\tau) &
\end{flalign}  
using the Continuous Time QMC
(CTQMC) within the Hybridization Expansion approach \cite{rmpqmc,PhysRevB.75.155113}. The self-consistent determination of
Eq.s (\ref{selfcon1}) and (\ref{selfcon2}) outputs $G_{\rm B/AB} (\omega)$ and the respective self-energies, which embody all
the physical information that we need.

Our main result is shown in Fig.\ref{nVsmu}, where we display the total density $n$ as a function of the
chemical potential $U/2- \mu$, shifted with respect to the half-filled case. As previously reported
\cite{PhysRevB.95.035113,PhysRevB.97.045108}, for $\mu=U/2$ the system is in a Mott insulator. This can be right away
seen by following the $n$ vs. $U-\mu/2$ behavior at the lowest temperatures $T$. It displays the horizontal plateau 
of electronic incompressibility at half-filling $n$ = 2, which is the hallmark of a
Mott insulating state. By varying $\mu$, we can dope particles and holes so to drive a transition to a 
metallic state. Here we consider the hole-doping case,
i.e. $n < 2$ (note that data would be symmetric for particle doping). At high temperatures
($T = 1/\beta > 1/20$), the transition to a metal takes place at small chemical potential $U/2-\mu$,
due to the fact that the thermal excited states occupying the Mott gap are available for adding holes.
At smaller temperatures, the metallic state emerges more steeply at the plateau edge. For $\beta = 50$, we can observe
that around $U/2-\mu \simeq 0.21$, the system is already deep in the metallic state and the slope of the
$n$ vs $U/2- \mu$ curve is markedly steep. This marks a diverging charge compressibility ($\propto dn/d\mu$) which is a
typical signature of the onset of a phase transition \cite{PhysRevLett.89.046401} (see SM \cite{sup1} for more details). Indeed, by further
reducing temperature ($T = 1/\beta  <  1/50$) a coexistence region between two distinct metallic solutions, one coming
from the insulator , which we shall show is a PG metal, and the other a FL coming from high doping, emerges 
and signals a first order metal-to-metal transition. 

\begin{figure}[h]
\begin{center}
\includegraphics[width=0.95\linewidth]{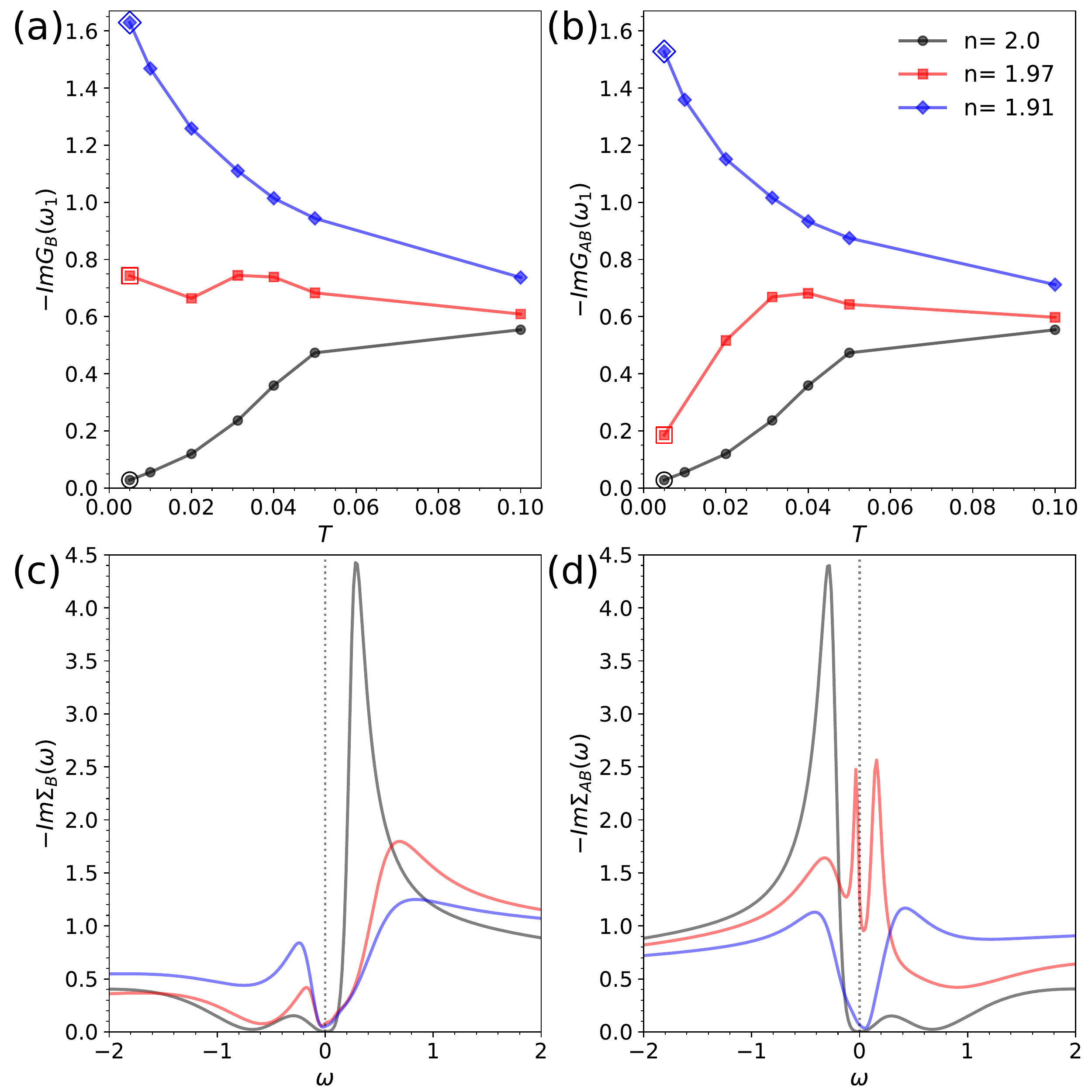}
\end{center}
\caption{Panels (a) and (b): Respectively, -Im$G_{B}(\omega_1)$ and -Im$G_{AB}( \omega_1)$ 
as a function of $T$ for densities $n$ = 2.0, 1.97, 1.91, where $\omega_1$ is the first Matsubara frequency.
Panels (c) and (d): Respectively, -Im$\Sigma_{B}(\omega)$ and -Im$\Sigma_{AB}(\omega)$ as a function 
of real frequency $\omega$ for the same densities and the lowest temperature $T = 1/\beta = 0.005$.}
\label{SG1Vsmu}
\end{figure}
To gain insight into the nature of these two qualitatively different metallic phases, we display in Fig.~\ref{SG1Vsmu}a,b the imaginary part of
the dimer Green's function, $G_{B/AB}(\omega_1)$, which directly outputs from the CTQMC-DMFT, evaluated
at the first Matsubara frequency
$\omega_1$, as a function of temperature. 
This is a useful quantity as it is a measure of the spectral intensity at low frequency.
Moreover, for $T \to 0$ it approaches the $D(E_F$), thus providing a convenient criterion to distinguish metal 
from insulator states.
A conventional FL metal is
characterized by the presence of a quasiparticle spectral weight at the Fermi level. 
Indeed, at doping $n= 1.91$
(blue-diamond line), we observe that both Im$G_{B/AB}(\omega_1)$ approach a similar finite value for $T \to 0$. 
The high doping phase is therefore a normal correlated FL. 
In contrast, in the insulating phase at half-filling  $n=2$, the low frequency spectral weight gets depleted 
(Im$G(\omega_1) \to 0$ as $T \to 0$)  as the Mott gap opens. 

The interesting behaviour appears for $n= 1.97$ in the low doping metallic phase. 
While the $B$ contribution displays metallic behavior 
as $G_{B}(\omega_1)$ extrapolates to a finite value for $T\to 0$,
the $AB$ contribution, in contrast, looses spectral intensity at low frequency. 
Indeed, Im$G_{AB}(\omega_1)$ extrapolates to
0, or to a small finite value, when $T \to 0$.
This anomalous state is an instance of orbital selectivity, which we call the PG metal
\cite{PhysRevB.80.064501}.

To better understand the physical meaning of these results, 
it is useful to display on the real-axis the imaginary part of the corresponding self-energy
$\Sigma_{B/AB}(\omega)= {{\cal G}_o}_{B/AB}^{-1}(\omega)- G_{B/AB}^{-1}(\omega)$.
The analytical continuation to the real axis of the CTQMC was obtained
with a standard maximum entropy method.
The self-energies are a measure of the correlation contribution
to the many-body system and reveal the qualitatively different nature of the metallic states (see Fig.\ref{SG1Vsmu}c,d). 
 We can first discuss the more conventional Mott-insulator and FL phases
at $n=2.0$ and $n=1.91$, respectively. 
Similarly to the familiar case of the single-site Mott insulator, where the correlation gap opens 
due to a pole in $\Sigma(\omega=0)$, in the DHM the Mott gap opens by the same mechanism.
More specifically, as seen in Fig.\ref{SG1Vsmu}c,d, the intra-dimer hopping splits the pole symmetrically in $\Sigma_{B/AB}$
at $\omega = \pm t_{perp}$, respectively \cite{PhysRevB.95.035113,PhysRevB.97.045108}.
The behavior of the self-energies sheds additional light on the physics of the DHM Mott insulator state.
In contrast to the single-site case, where the localized electrons are incoherent spin-1/2, here they are in dimers 
that form a liquid of singlets, as envisioned by Anderson \cite{ANDERSON1196}. 
Thus, the electrons spend most of their time within a well defined quantum state, which gives them a long-lived 
quasiparticle-like character, which is reflected in the behavior of Im$\Sigma \to 0$ 
at $\omega \to 0$. 

Upon doping, the peak feature is strongly reduced, the Mott gap closes, and the relevant energy scale is close to $\omega=0$. In a metal,
the FL prescription requires that Im$\Sigma\sim \omega^2$.
This behavior is observed in the FL phase ($n=1.91$ blue line), both in the Im$\Sigma_{B/AB}$,
in agreement with the behavior of Im$G_{AB}$ discussed in Fig.\ref{SG1Vsmu}a,b.
On the other hand, in the PG metal ($n=1.97$ red line), while the Im$\Sigma_B$ still displays a FL behavior
at low frequency,
the Im$\Sigma_{AB}$ completely breaks the FL, displaying a peak-like behavior at $\omega=0$, 
reminiscent of previous results on a slightly doped Mott-Hubbard insulator in 2D \cite{PhysRevLett.120.067002,PhysRevB.80.064501}.

The PG-metal appears to be smoothly connected to the Mott insulating state, as shown by the smooth evolution
of the plateau in $n$ vs. $\mu$ seen in Fig.\ref{nVsmu} \cite{PhysRevLett.104.226402}. The origin of the 
PG insulating component can be traced back to the Mott insulating state, which in the DHM is sharply distinct from the
paramagnetic Mott insulator of single-site DMFT \cite{PhysRevB.95.035113,PhysRevB.97.045108}.
In the DHM the physical properties are dominated by singlet formation (via correlation enhancement of the intra-dimer
hopping $t_{\perp}$ \cite{PhysRevB.95.035113}), which is a non-local magnetic interaction that competes with the 
on-site Kondo mechanism. 
To show this connection,
in Fig.\ref{nVsmu} we display pie charts showing the singlet contribution to the wave-function projection on the
dimer sites. In the Mott insulator the singlet contribution is 91\%, and it is still a largely majoritarian 85\% in the PG phase, once
holes are introduced into the systems. This must be contrasted with FL phase, where the singlet appear to have a 50\% contribution,
comparable with that of other components (see SM \cite{sup1} for details about this analysis).
The PG phase can be qualitatively viewed as a soup of singlets and strongly damped (as described by -Im$\Sigma_{B}(\omega)$ 
in Fig.\ref{SG1Vsmu}c) FL-like hole carriers.

 \begin{figure}[!tb]
\begin{center}
\includegraphics[width=.99\linewidth]{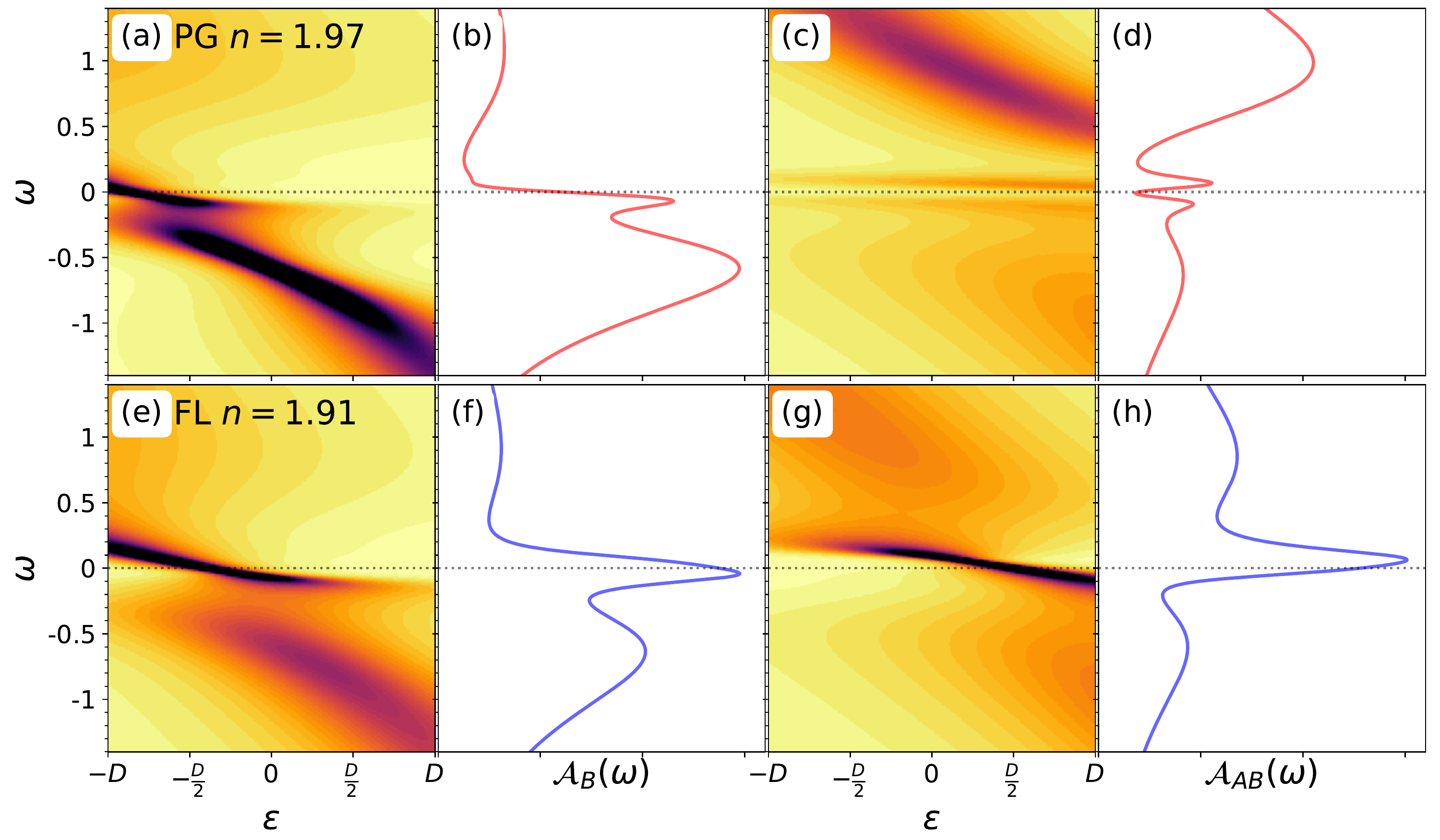}
\end{center}
\caption{Electronic dispersion $A_{B/AB}(\omega,\epsilon)$ and local density of states 
$A_{B/AB}(\omega)= \int d\epsilon  A_{B/AB}(\omega,\epsilon)$ ($B$ first and second column
and $AB$ third and fourth). Top line has the PG-metal and the bottom line the FL-metal.}
 \label{transport}
\end{figure}
 \begin{figure}[!!tb]
   \begin{center}
     \includegraphics[width=.65\linewidth]{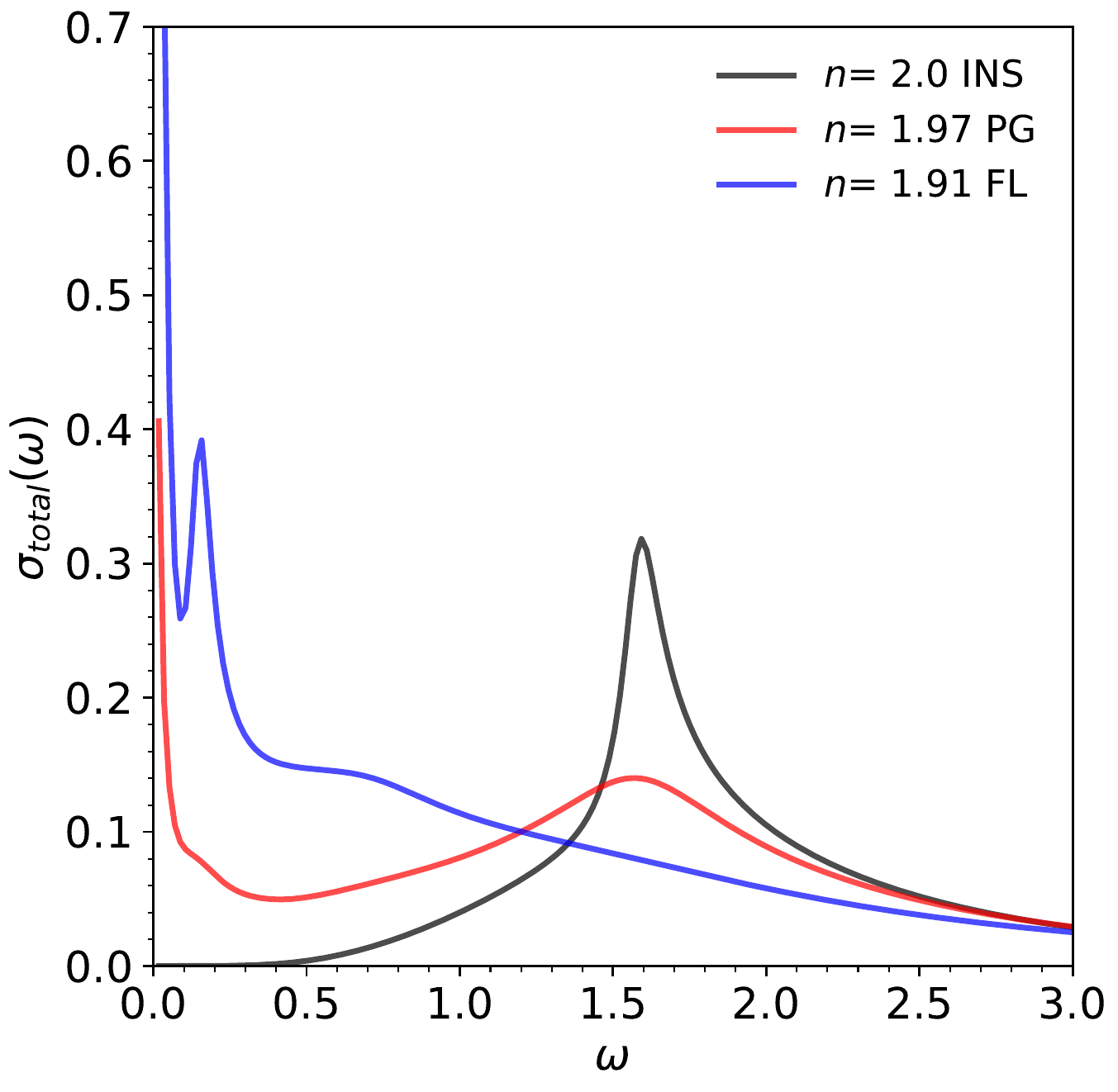}
   \end{center}
   \caption{Optical conductivity in the insulating ($n=2.0$) in black, PG ($n=1.97$) in red and FL phases ($n=1.91$) in blue.}\label{optcond}
 \end{figure}

 We argued before that the DHM PG-metal exhibits orbital selectivity. We shall now show that, interestingly,
that state bears resemblance to the results from cluster extensions of the Hubbard Model in 2D
\cite{kyung,civelliBreakup,Macridin2006}, where the PG-state was argued to emerge from 
a ``momentum-selective'' Mott transition
\cite{michelPRB,gullPRB,PhysRevB.84.075161,sshtRHO,ssht}.
%
 This can be seen from the $\epsilon$-resolved spectral functions $A(\omega,\epsilon)$, where single particle
 energy $\epsilon$ plays a similar role as the lattice momentum \cite{RMP}, since
 $A_{B/AB}(\omega,\epsilon)= \rm{Im}[1/(\omega+ \mu-\epsilon- \Sigma_{B/AB}(\omega))]$.
 The results are shown in Fig.~\ref{transport}, 
 which we obtain by analytic continuation. 
 At small doping ($n=1.97$), we obtain the PG metal solution. In the B component, a coherent and dispersive narrow band appears
 at the Fermi level, which produces a narrow FL peak in the respective DOS.
 In stark contrast, the AB component develops two parallel and weakly-dispersive features around the Fermi level,
 which produce the PG-like dip in the DOS. Therefore, the PG metal possesses at the same time FL and non-FL character,
 qualitatively similar to the cluster extension and in strong analogy with the physics of cuprates. 
 On the other hand, at higher doping ($n=1.91$), the orbital selectivity disappears at low frequency. Both components 
 display similarly dispersive coherent quasiparticle bands and quasiparticle peaks in the respective DOS,
 as expected in a conventional FL. 
 
 Finally, we obtained the optical conductivity that we show in Fig.\ref{optcond} for the Mott insulator along with the 
 PG and the FL metal states \cite{PhysRevB.95.035113}.
 The Mott insulator has as expected a wide gap (black line $n=2$). In contrast, at high doping (blue line $n=1.94$) 
 the optical response, as expected in a correlated metal with quasiparticles, has a narrow Drude peak at low frequency,
 whose spectral intensity denotes the effective carrier density. In addition, there is a mid-infrared contribution that has been
 previously identified as originated in the intra-dimer hopping \cite{PhysRevB.95.035113}.  
 Interestingly, in between these two states we find the optical response of the PG-metal (red line $n=1.97$), 
 which combines features of both.
 At low frequency, it shows a narrow and relatively small Drude peak that can be traced to the FL-metal of the $B$ component.
 This contribution coexists with significant mid-infrared intensity, which is featureless as it is associated to the incoherent 
 PG state realized in the $AB$ component. It is interesting to note that this optical conductivity response bears strong 
 resemblance with the one observed in the PG phase of cuprates \cite{RevModPhys.77.721,PhysRevB.47.8233}.

 To conclude, we obtained the numerical quantum Monte Carlo solution on the Dimer Hubbard Model within DMFT, 
 which is the exact solution of the model in the limit of large dimensions. This model is arguably the minimal Hubbard-like
 model that embodies on equal footing on-site correlations and non-local magnetic interactions.
 Our main result was to establish the existence of the doping-driven
 first order metal-to-metal transition between a pseudogap metal and a Fermi Liquid one.
 Moreover, the scenario that emerges from our findings help to clarify an important
 on-going debate. 
 It provides a concrete realization and an explicit rational for the origin of several unconventional properties  
 such as bad metal behavior, the pseudogap, orbital selective Mott transitions and enhanced compressibilities
 \cite{PhysRevB.84.075161,sshtRHO,ssht,fratino2016organizing,PhysRevB.93.245147}, which are relevant to 
 doped Mott insulators and correlated quantum materials in 
 general \cite{RevModPhys.87.457, Efetov2013, PhysRevLett.64.475}.
 
 
 
SB, MC and MR acknowledge support from the French ANR ``MoMA'' project ANR-19-CE30-0020. 
LF is supported by the UCSD-CNRS collaboration Quantum Materials for Energy Efficient Neuromorphic Computing, an
Energy Frontier Research Center funded by the US Department of Energy,
Office of Science, Basic Energy Sciences under award DE-SC0019273.
 
 \bibliographystyle{unsrtnat}
 \bibliography{bibliography_3}{}
 

\end{document}